\documentclass[twocolumn,showpacs,preprintnumbers,amsmath,amssymb]{revtex4}

\usepackage{graphicx}
\usepackage{dcolumn}
\usepackage{bm}

\begin{document}

\preprint{APS/123-QED}

\title{Onsager reciprocity in premelting solids}

\author{ S. S. L. Peppin$^1$}

\author{M. Spannuth$^1$}

\author{J. S. Wettlaufer$^{1,2}$}%

\affiliation{%
$^1$Department of Geology and Geophysics and $^2$Department of Physics\\
Yale University, New Haven, CT, 06520
}%

\date{\today}

\begin{abstract}
The diffusive motion of colloidal particles dispersed in a
premelting solid is analyzed within the framework of irreversible
thermodynamics. We determine the mass diffusion coefficient, thermal
diffusion coefficient and Soret coefficient of the particles in the
dilute limit, and find good agreement with experimental data. In
contrast to liquid suspensions, the unique nature of premelting
solids allows us to derive an expression for the Dufour coefficient
and independently verify the Onsager reciprocal relation coupling
diffusion to the flow of heat.

\end{abstract}

\maketitle

The melting of any material is normally initiated at one of its free
surfaces at temperatures below the bulk melting temperature, $T_m$,
by the formation of a thin liquid -- {\em interfacially premelted}
-- film. This surface phase transition has been observed at the interfaces of
solid rare gases, quantum solids, metals, semiconductors and
molecular solids including ice, allowing the liquid phase to persist
in the solid region of the bulk phase diagram \cite{Dash:06}. 
When $T \approx T_m$  the premelted film is thicker 
than the correlation lengths of the solid-liquid interfaces and hence the total free energy of the (planar-planar) system is represented as a linear combination of bulk and interfacial terms (this latter contains both the interfacial energy and the long ranged volume-volume interactions which are also forces/area in such a system).  The total free energy of a {\em curved} interfacially melted system includes in addition the Gibbs-Thomson effect, which, while contributing to the premelted film thickness $d$, one can prove produces no net thermodynamic force over a closed surface \cite{Rempel:01}.  Nonetheless, when a
premelted film forms around a foreign particle within a subfreezing
solid, the particle can migrate under the influence of a temperature
gradient, which produces a thermomolecular pressure gradient, a phenomenon referred to as regelation or thermodynamic
buoyancy \cite{Rempel:01,Wettlaufer:06}. Here we analyze this and
related phenomena within the framework of irreversible
thermodynamics and demonstrate that in addition to motion under the
influence of a temperature gradient, the particle can undergo
constrained Brownian motion owing to thermal fluctuations in the
premelted film. We determine the diffusivity in the dilute limit and
relate it to the Stokes-Einstein diffusivity of particles in the
bulk melt. Furthermore, the motion of a particle is shown to induce
a reciprocal effect -- a heat flux due to the release and absorption
of latent heat on opposite sides of the particle. We show that this
effect is described by the Onsager relation coupling mass diffusion
to heat flux.

\begin{figure}
\centering
\includegraphics[width=240pt]{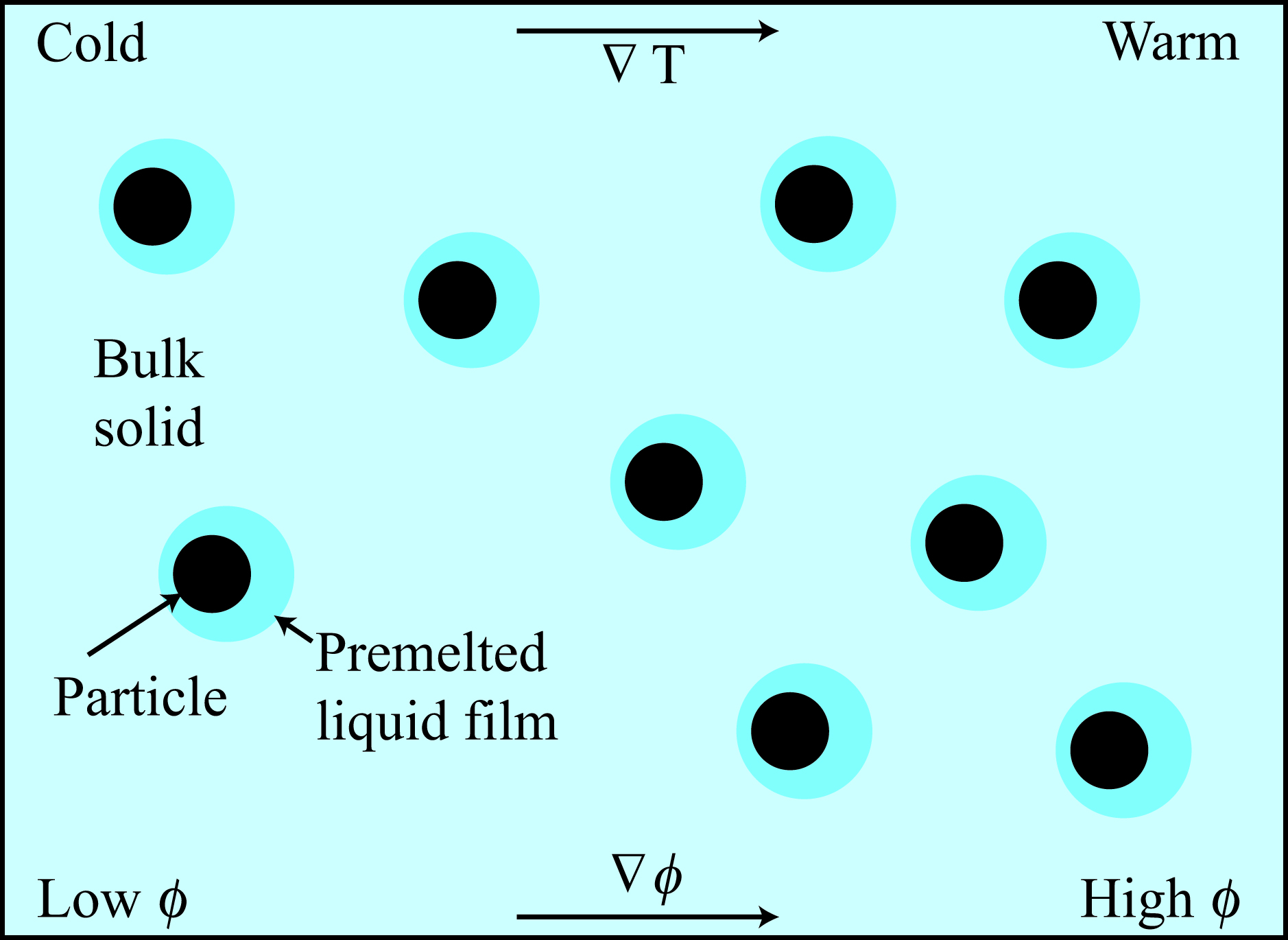}
\caption[] { Schematic diagram of a premelting solid containing
colloidal inclusions at local volume fraction $\phi$.
 } \label{fig:schem_dsc}
\end{figure}

We consider a suspension of spherical particles of radius $R$
randomly distributed in a bulk solid. For temperatures $T$ near but
less than $T_m$, each particle is surrounded by a premelted film of
thickness $d$ that facilitates both the particles' constrained
Brownian diffusion and their directed motion parallel to temperature
gradients (figure \ref{fig:schem_dsc}). From linear irreversible
thermodynamics \cite{deGroot:62, Onsager:31b}, the equations
describing the flux of particles and heat are (see Appendix)
\begin{equation} 
\label{eq:mss_flx}
    \textbf{J} = -D\nabla\phi + (1-\phi)\phi D_T\nabla T,
\end{equation}
\begin{equation} 
\label{eq:ht_flx}
    \textbf{q} = -k\nabla T + (1-\phi)\Pi_\phi T_mD_q\nabla\phi,
\end{equation}
where $\textbf{J}=\phi \textbf{V}$ is the particle volume flux, with
$\phi$ the volume fraction of particles and $\textbf{V}$ the local
average particle velocity. In equation (\ref{eq:ht_flx})
$\textbf{q}$ is the heat flux, $k$ is the thermal conductivity and
$\Pi_\phi$ is the derivative of the osmotic pressure of the
particles with respect to volume fraction. The coefficients $D$,
$D_T$ and $D_q$ are, respectively, the mass diffusion coefficient,
thermal diffusion coefficient and Dufour coefficient. Here, in the
dilute limit, we determine these quantities as functions of the
undercooling $\Delta T = T_m-T$.

The second term on the right hand side of equation
(\ref{eq:mss_flx}) accounts for the Soret effect, in which a
temperature gradient generates motion of the particles. We define
$D_T$ to be positive when the particles migrate towards their warm
sides. Whereas in liquid suspensions the Soret effect is a complex
phenomenon, with the Soret coefficient $S_T=D_T/D$ often changing
sign as the temperature increases \cite{Piazza:04,Braibanti:08}, in
premelting solids the particles always appear to migrate towards
higher temperatures. Furthermore, the Soret coefficient is several
orders of magnitude {\em larger} than in liquid systems.

The second term in equation (\ref{eq:ht_flx}) represents the Dufour
effect, in which a concentration gradient gives rise to a flow of
heat. Using the principle of microscopic reversibility, Onsager
proved that the Dufour coefficient $D_q$ is equivalent to the
thermal diffusion coefficient $D_T$ \cite{Onsager:31b,deGroot:62}.
There has, however, been significant controversy over the range of
validity of the Onsager relations \cite{Coleman:60,Truesdell:69} and
ongoing attempts to prove them in particular cases
\cite{Kirkwood:60,Mullins:80,Elliott:00}. We show that, in the
dilute limit, so long as the temperature gradient satisfies linear
irreversible thermodynamics and is constant on the scale of 
the particle, the Onsager relation is satisfied.

Colloidal particles dispersed throughout a solid near its melting
temperature undergo random motion owing to thermal fluctuations in
their premelted films. Their random walk is driven by the
thermodynamic force
    \begin{equation} \label{eq:therm_frc}
    \textbf{F}_B = -\frac{v_p}{\phi}\nabla\Pi,
    \end{equation}
where $\Pi$ is the osmotic pressure of the particles and $v_p =
\frac{4}{3}\pi R^3$ is the particle volume \cite{Einstein:56}. For
temperatures near $T_m$ in the dilute limit the osmotic pressure is
given by the van't Hoff equation
    \begin{equation} \label{eq:def_osmP}
    \Pi = \frac{\phi}{v_p}k_bT_m,
    \end{equation}
where $k_b$ is Boltzmann's constant.

Owing to viscous flow in the premelted films, the particle velocity
$\textbf{V}$ is accompanied by a net lubrication force given by
\begin{equation} \label{eq:visc_frc}
    \textbf{F}_\eta = -\upsilon_p\frac{\eta}{\textit{\textsf K}}\textbf{V},
 \end{equation}
where ${\textit{\textsf K}}=d^3/6R$ is a permeability coefficient
\cite{Rempel:04} and $\eta$ is the dynamic viscosity of the
premelted liquid film.

In mechanical and thermal equilibrium, the total force ${\bf F}_\eta+{\bf F}_B \equiv 0$, giving
    \begin{equation} \label{eq:diff_eq2}
    \phi\textbf{V} =
    -\frac{d^3k_bT_m}{8\pi\eta R^4}\nabla\phi,
    \end{equation}
Comparing this result with equation (\ref{eq:mss_flx}) yields
  \begin{equation}
  \label{eq:SE_diff}
    D= \left(\frac{3d^3}{4R^3}\right)D_{0} ~~~{\textrm{where}} ~~~ D_{0} = \frac{k_bT_m}{6\pi\eta
    R}.
    \end{equation}
Hence we see that the premelting-controlled Brownian diffusivity $D$
differs from the Stokes-Einstein diffusivity $D_{0}$ of particles in
bulk liquid by the factor $3d^3/4R^3$. Preliminary studies using
X-ray photon correlation spectroscopy show promise in testing
equation (\ref{eq:SE_diff}) \cite{Spannuth:07}.

\begin{figure}
\centering
\includegraphics[width=240pt]{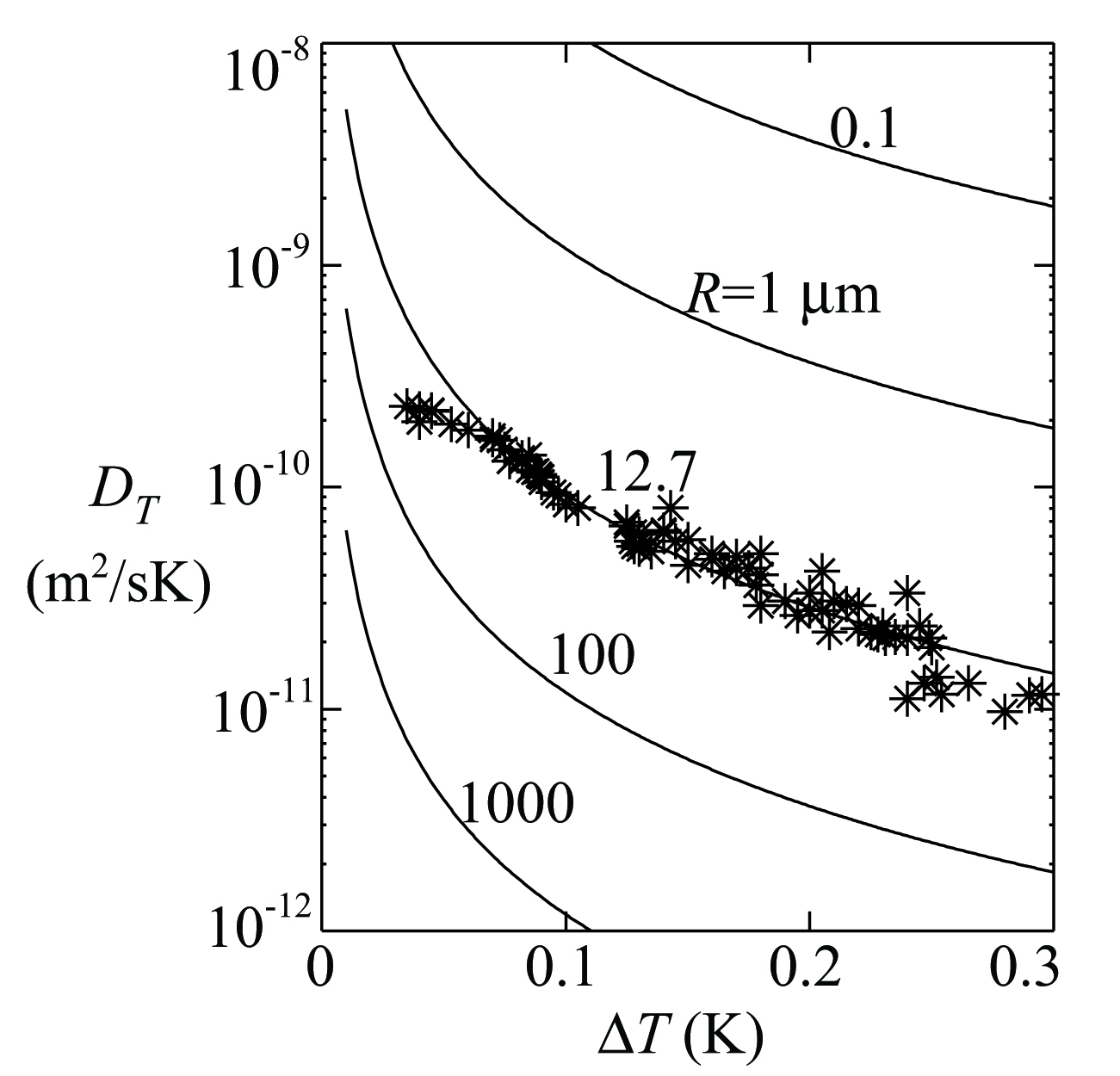}
\caption[] { Thermal diffusion coefficient of silica particles in
ice. The data (stars) are from Romkens and Miller \cite{Romkens:73},
while the theoretical curves are from equation
(\ref{eq:thmm_dff_cf}) using $d$ obtained from the averaged
experimental measurements of Ishizaki \emph{et al.}
\cite{Ishizaki:96} and Engemann \emph{et al.} \cite{Engemann:04} at
each $\Delta T$. The properties of ice and water at $T_m$ are:
$\rho_s = 920\,$kg/m$^3$, $\eta=0.0017\,$N\,s/m$^2$, $q_m =
3.34\times10^5\,$J/kg.} \label{fig:part_spd}.

\end{figure}

In the presence of a temperature gradient variations in the
thickness of the premelted film over the surface of an isolated
particle lead to a thermomolecular pressure gradient force causing
the particle to move \cite{Dash:06, Wettlaufer:06, Rempel:01}. By integrating the fluid
pressure over the surface of a particle this force has been obtained
as
    \begin{equation} \label{eq:thmm_frc}
    \textbf{F}_T = \upsilon_p{\cal P}_T\nabla T,
    \end{equation}
where ${\cal P}_T \equiv\rho_sq_m/T_m$ is the thermomolecular pressure
coefficient characterizing the magnitude of the intermolecular
forces responsible for the premelted film \cite{Dash:06}, and
$\rho_s$ and $q_m$ are the mass density and latent heat of fusion,
respectively, of the bulk solid \cite{Rempel:01}. (We express Eq. (\ref{eq:thmm_frc})
differently than in \cite{Rempel:01} for consistency with the
present extension of that work.) In mechanical
equilibrium the thermomolecular force is balanced by the viscous
force ${\bf F}_\eta$ leading to the relation
    \begin{equation} \label{eq:th_bf}
    \textbf{V} = \frac{\textit{\textsf K}}{\eta}{\cal P}_T\nabla T,
    \qquad (\nabla\phi = 0).
    \end{equation}
Comparing (\ref{eq:mss_flx}) with (\ref{eq:th_bf}) in the limit
$\phi\rightarrow 0$ yields an expression for the thermal diffusion
coefficient
    \begin{equation} \label{eq:thmm_dff_cf}
    D_T = \frac{\textit{\textsf K}}{\eta}{\cal P}_T =
    \frac{d^3\rho_s q_m}{6R\eta T_m}.
     \end{equation}
This relation can be tested experimentally using the results of a
study in which a 12.7\,$\mu$m glass bead encased in ice was moved by
imposition of a linear temperature gradient \cite{Romkens:73}.
Figure \ref{fig:part_spd} shows experimental measurements (stars) of
$D_T=\textbf{V}/\nabla T$ as a function of the undercooling. Using
measurements of $d=d(\Delta T)$ \cite{Ishizaki:96,Engemann:04} in
equation (\ref{eq:thmm_dff_cf}) we can predict $D_T(\Delta T)$ which
is shown by the solid lines. Clearly the agreement between theory
and experiment is good. The deviation at small $\Delta T$ may be due
to an overestimation of the film thickness by Ishizaki \textit{et
al.} \cite{Ishizaki:96} owing to curvature effects in their system,
which become important as $\Delta T$ approaches 0.

The Soret coefficient $S_T$ is defined as
    \begin{equation} \label{eq:Soretdef}
    S_T \equiv \frac{1}{\phi(1-\phi)}\frac{\nabla\phi}{\nabla
    T}\Big{|}_{\textbf{J}=0}.
    \end{equation}
Combining (\ref{eq:mss_flx}), (\ref{eq:SE_diff}) and
(\ref{eq:thmm_dff_cf}) with (\ref{eq:Soretdef}) gives
    \begin{equation} \label{eq:Soret}
    S_T =\frac{D_T}{D}=\frac{{\cal P}_T}{{\cal E}_T},
    \end{equation}
in terms of fundamental quantities, where here we emphasize that
${{\cal E}_T}\equiv{k_bT_m}/{v_p}$ is the ensemble averaged kinetic
energy/volume (thermal kinetic energy) of a Brownian particle. Thus
in premelting solids the Soret coefficient represents the ratio of
the thermomolecular pressure coefficient to the thermal kinetic
energy of a Brownian particle and is predicted to be independent of
premelted film thickness and undercooling. For micron sized
particles in ice equation (\ref{eq:Soret}) yields $S_T\sim
10^8\,K^{-1}$ thereby illustrating the dominance of the
thermomolecular force over the Brownian force in premelting solids.
As a comparison, for $\mu$m sized particles in liquids $S_T\sim
10^{-2}\,K^{-1}$ \cite{Piazza:04}. It is hoped that these forms of
$S_T$ provide an ideal experimental target.  One promising method
for obtaining both $D$ and $D_T$ simultaneously is thermal diffusion
forced Rayleigh scattering \cite{Kohler:93}.

When foreign particles of volume $v_p$ are inserted into a
premelting host material the heat of fusion absorbed by the solid
per unit volume of particles is $\rho_sq_m$. Whereas, when particles
move through an initially isothermal solid they melt material ahead,
absorbing latent heat, while freezing solid behind, releasing heat.
The volumetric heat of transport characterizing this phenomenon is
defined by
    \begin{equation} \label{eq:def_ht_tns}
    Q_v^*\equiv \frac{\textbf{q}}{\textbf{J}}\Big|_{\nabla T =0} =
    -{(1-\phi)\Pi_\phi T_m}\frac{D_q}{D},
    \end{equation}
where the second equality follows from (\ref{eq:mss_flx}) and
(\ref{eq:ht_flx}). Hence, physically, $Q_v^*$ represents the
sensible heat released by an arbitrarily isolated part of the system
per unit volume of particles diffusing into it (e.g.,
\cite{Eastman:28,Bearman:58}) and we obtain $Q_v^* = -\rho_sq_m$.

We verify the reciprocal relation coupling mass diffusion to heat flux in
our system by inserting (\ref{eq:def_osmP}), (\ref{eq:SE_diff}),
(\ref{eq:thmm_dff_cf}) and $Q_v^*=-\rho_sq_m$ into equation
(\ref{eq:def_ht_tns}) and taking the limit $\phi\rightarrow0$ to
obtain
    \begin{equation} \label{eq:ORR}
    D_q = D_T.
    \end{equation}
If the Onsager relation (\ref{eq:ORR}) holds also at higher
concentrations, (\ref{eq:Soretdef}), (\ref{eq:def_ht_tns}) and
(\ref{eq:ORR}) can be used to obtain an expression for the
concentration dependence of the Soret coefficient:
    \begin{equation} \label{eq:ST_new}
    S_T = -\frac{Q_v^*}{(1-\phi)\Pi_\phi T_m}.
    \end{equation}
The Onsager relation will break down  when the temperature gradient
varies significantly over the surface of a particle because,  as
noted by Rempel \emph{et al.} \cite{Rempel:01}, under these
circumstances the thermomolecular pressure gradient will vary 
over the surface of the particle and hence the expression (\ref{eq:thmm_frc}) for the
thermomolecular force will no longer hold.  Such a constraint on $\nabla T$ is independent
of the issue of the validity of linear irreversible thermodynamics as embodied in 
equations (\ref{eq:mss_flx}) and (\ref{eq:ht_flx}).  This is because while  $\nabla T$ may be sufficiently
small for these relations to be satisfied it may still vary on the particle scale, say, for large particles.  Nonetheless, for most experimental situations envisaged the temperature gradient is sufficiently small that the Onsager relation holds.  

We have shown the complimentary roles that interfacial premelting
plays in both the regelation and Brownian motion of colloidal
particles in bulk solids. In the dilute limit, within the framework
of linear irreversible thermodynamics,  the Brownian diffusivity and
Soret coefficient have been determined. Furthermore, in premelting
solids the heat of transport can be explicitly calculated, leading
to a verification of the Onsager relation coupling mass diffusion to
the flow of heat.

\begin{acknowledgments}
This research was supported by the Department of Energy (DE-FG02-05ER15741) and by
the U.S. National Science Foundation (OPP0440841).
M.S. acknowledges support from a U.S. National Science Foundation Graduate Research Fellowship.
\end{acknowledgments}

\appendix*
\section{Mass and heat fluxes}
Here we show that equations (\ref{eq:mss_flx}) and (\ref{eq:ht_flx})
can be derived from the irreversible thermodynamic treatment of the flux
of mass and heat in a two-component hydrostatic mixture (cf., 
equations (XI.226) and (XI.227) of \cite{deGroot:62}), viz.,  
    \begin{equation} \label{eq:DMmss_flx}
    \textbf{J}_1 = -\rho D\nabla c_1 - \rho c_1c_2 D_T\nabla T,
    \end{equation}
    \begin{equation} \label{eq:DMht_flx}
    \textbf{J}'_q = -k\nabla T - \rho_1 \left(\frac{\partial \mu_1}{\partial
c_1}\right)_{T,P}TD_q\nabla c_1.  
    \end{equation}
Here $\textbf{J}_1=\rho_1(\textbf{v}_1-\textbf{v})$ is the mass flux of
component $1$ relative to the barycentric velocity
$\textbf{v}=c_1\textbf{v}_1 + c_2\textbf{v}_2$, $c_j =
\rho_j/\rho$ is the mass fraction of $j$ which has a partial
mass density $\rho_j$, $\rho = \rho_1 + \rho_2$ is the mixture
density, and $\textbf{v}_j$ is the velocity of $j$ with respect to the
laboratory frame.  The so-called reduced heat flux is 
$\textbf{J}'_q$, where  $T$ is absolute temperature, $P$ is the mixture
pressure and $\mu_j$ is the chemical potential of $j$. By definition
    \begin{equation} \label{eq:cks}
    c_1 + c_2 = 1 \qquad \textrm{and} \qquad \textbf{J}_1 + \textbf{J}_2 = 0.
    \end{equation}
With (\ref{eq:cks}) equations (\ref{eq:DMmss_flx}) and
(\ref{eq:DMht_flx}) can be written as
    \begin{equation} \label{eq:DMmss_flx2}
    \textbf{J}_2 = -\rho D\nabla c_2 + \rho c_1c_2 D_T\nabla T,
    \end{equation}
    \begin{equation} \label{eq:DMht_flx2}
    \textbf{J}'_q = -k\nabla T - \rho_1 \left(\frac{\partial \mu_1}{\partial c_2}\right)_{T,P}TD_q\nabla
    c_2.
    \end{equation}

It is convenient to write equations (\ref{eq:DMmss_flx2}) and
(\ref{eq:DMht_flx2}) in terms of the partial density $\rho_2$.  To achieve this we use
the thermodynamic identity (equation XI.99 of \cite{deGroot:62})
    \begin{equation}    \label{eq:theid1}
    \left(\frac{\partial \rho_2}{\partial c_2}\right)_{T,P} =
    \rho^2\nu_1,
    \end{equation}
where $\nu_1$ is the partial specific volume of component 1,
approximately equal to
    \begin{equation}    \label{eq:pspecvl}
    \nu_1 = 1/\rho_s,
    \end{equation}
where $\rho_s$ is the density of pure component 1 at $T$ and $P$.
Additionally, we make use of the fact that the mass flux can be written in terms of the volume average
velocity $\textbf{v}^0 = \rho_1\nu_1\textbf{v}_1 +
\rho_2\nu_2\textbf{v}_2$ using the thermodynamic identity
$\rho_1\nu_1 + \rho_2\nu_2 = 1$ to obtain
    \begin{equation}    \label{eq:flx_trns}
    \textbf{J}_2^0 = \rho\nu_1\textbf{J}_2 
    \end{equation}
where $\textbf{J}_2^0 = \rho_2(\textbf{v}_2 - \textbf{v}^0)$.

With (\ref{eq:theid1}), we rewrite (\ref{eq:DMmss_flx2}) in the form of (\ref{eq:flx_trns}) and this and (\ref{eq:DMht_flx2}) become
    \begin{equation} \label{eq:DMmss_flx3}
    \textbf{J}_2^0 = - D\nabla \rho_2 + \nu_1\rho_1\rho_2 D_T\nabla T,
    \end{equation}
    \begin{equation} \label{eq:DMht_flx3}
    \textbf{J}'_q = -k\nabla T - \rho_1 \left(\frac{\partial \mu_1}{\partial\rho_2}\right)_{T,P}TD_q\nabla
    \rho_2.
    \end{equation}
For a dispersion of hard spherical particles $\nu_2 = 1/\rho_p$,
where $\rho_p$ is the constant mass density of an individual
particle, in which case (\ref{eq:DMmss_flx3}) and
(\ref{eq:DMht_flx3}) can be written as
    \begin{equation} \label{eq:DMmss_flx4}
    \textbf{J} = - D\nabla \phi + \phi(1-\phi) D_T\nabla T,
    \end{equation}
    \begin{equation} \label{eq:DMht_flx4}
    \textbf{q} = -k\nabla T - \rho_1 \left(\frac{\partial \mu_1}{\partial\phi}\right)_{T,P}TD_q\nabla
    \phi,
    \end{equation}
where $\textbf{J} = \phi(\textbf{v}_2 - \textbf{v}^0)$ is the
particle volume flux, $\phi = \rho_2/\rho_p$ is the volume fraction
of particles and $\textbf{q} = \textbf{J}'_q$. 
As shown by de Groot and Mazur,
in closed systems in which $\nu_1$ and $\nu_2$ are constant, the
volume average velocity $\textbf{v}_0$ is zero, in which case
(\ref{eq:DMmss_flx4}) reduces to (\ref{eq:mss_flx}).
Inserting the
thermodynamic relation $(\partial \mu_1/\partial\phi)_{T_m,P_m} =
-(1/\rho_s)\Pi_\phi$ into (\ref{eq:DMht_flx4}) gives equation
(\ref{eq:ht_flx}), where $1-\phi = \rho_1/\rho_s$ and $\Pi_\phi =
(\partial\Pi/\partial\phi)_{T_m,P_m}$ at a reference state of bulk coexistence.

\bibliography{premelt}

\end{document}